\documentclass[aps,prl,showpacs,notitlepage,nofootinbib,superscriptaddress,floatfix,showkeys,twocolumn]{revtex4-2}

\usepackage{blindtext}
\usepackage{hyperref}
\usepackage{amsmath,amssymb}
\usepackage{float}
\usepackage{microtype}
\usepackage{marginnote}
\usepackage{graphicx}
\usepackage{bm}
\usepackage{latexsym}
\usepackage{epsfig}
\usepackage{psfrag}
\usepackage{color}
\usepackage[dvipsnames]{xcolor}
\usepackage{subfigure}
\usepackage{graphicx} 
\usepackage{amssymb}
\usepackage{amsmath}
\usepackage{bm}
\usepackage{latexsym}
\usepackage{epsfig}
\usepackage{psfrag}
\usepackage[normalem]{ulem}
\usepackage{textcomp}
\usepackage{color}
\usepackage{pstricks}
\usepackage[utf8]{inputenc}
\usepackage{comment}
\usepackage{braket}
\usepackage{mathrsfs}
\usepackage{cleveref}
\usepackage{commath}

\def\bea{\begin{eqnarray}}
\def\eea{\end{eqnarray}}
\def\nn{\nonumber}

\def\f{\frac}
\def\l{\left}
\def\r{\right}
\def\d{{\rm d}}
\def\Mpl{M_{_{\mathrm{Pl}}}}
\def\mpcinv{{\rm Mpc}^{-1}}

\def\cR{{\mathcal R}}
\def\cRk{\mathcal{R}_{\bm{k}}}
\def\vx{{\bm{x}}}
\def\vk{{\bm{k}}}
\def\ps{\mathcal{P}_{_{\mathrm{S}}}}
\def\sps{\mathscr{P}_{_{\rm S}}}

\def\barpow1d{\mathcal{P}^{1\rm D}_{\mathrm{b}}}

\def\fnl{f_{_{\rm NL}}}

\newcommand{\e}{\text{e}}

\newcommand{\Ro}{\hat{\mathcal{R}}}

\newcommand{\IR}{\text{irr}}
\def\Dd{\delta^{(3)}}

\allowdisplaybreaks

\begin{document}

\title{Cosmological consequences of statistical inhomogeneity}
\author{H.~V.~Ragavendra}
\email{E-mail: ragavendra.hv@pd.infn.it}
\affiliation{Dipartimento di Fisica e Astronomia “Galileo Galilei”, Università degli Studi di Padova, Via Marzolo 8, I-35131 Padova, Italy}
\affiliation{Istituto Nazionale di Fisica Nucleare, Sezione di Padova, Via Marzolo 8, I-35131 Padova, Italy}
\author{Dipayan Mukherjee}
\email{E-mail: dipayanmkh@gmail.com}
\author{Shiv K.~Sethi}
\email{E-mail: sethi@rri.res.in} 
\affiliation{Raman Research Institute, C.~V.~Raman Avenue, Sadashivanagar, Bengaluru 560080, India}

\begin{abstract}
A space-dependent mean for cosmological perturbations negates the ansatz of statistical homogeneity and isotropy, and hence ergodicity. In this work, we construct such a primordial mean of scalar perturbations from an alternative quantum initial state (coherent state) and examine the associated power and bi-spectra. A multitude of cosmological tests based on these spectra are discussed. We find that current cosmological data doesn't favor a primordial mean over large scales and strong constraints arise from the limit on bispectrum from Planck data. At small scales, this hypothesis can be tested by future observables such as $\mu$-distortion of CMB.
\end{abstract}

\maketitle
\noindent
\underline{\textit{Introduction}}--- Inflation provides a natural
mechanism to impose initial conditions for cosmological
perturbations
~\cite{Linde:1990flp,Liddle:1994dx,Liddle_Lyth_2000,Riotto:2002yw,Martin:2003bt,Baumann2009,Sriramkumar:2009kg,Sriramkumar:2012mik,baumann2022}.
The quantum fluctuations generated during inflation evolve to become
the anisotropies in the cosmic microwave background (CMB) and large
scale structure in the current universe. The structure of the
N-point correlations of quantum fluctuations is inherited by
cosmological observables. In the simplest inflationary models, the
fluctuations are statistically homogeneous, statistically isotropic,
and Gaussian. A deviation from these generic properties could point
to more complicated inflationary models or quantum initial
conditions. The current data has not revealed any deviation from
either Gaussianity or statistical isotropy, even though there are
reports of related
anomalies~\cite{Planck:2018jri,Planck:2019kim,eBOSS:2021jbt,Schwarz_2016,
  Planck:2019evm, Vielva_2010, Inoue_2006,Vielva_2004, Akrami_2014,
  Adhikari_2014, Gordon_2007, Hansen_2004, Land_2005,Schwarz_2004,
  de_Oliveira_Costa_2004, Kothari_2016, Rath_2015, Carroll:2008br,
  Souradeep_2006,Bartolo_2012, Gordon_2007,Prunet:2004zy,Aluri:2022hzs}.

In the usual case, the primordial fluctuations are assumed to have
evolved from the Bunch-Davies vacuum
state~\cite{Bunch:1978yq,Martin:2004um,Sriramkumar:2012mik}.
However, it is not a unique choice~\cite{Armendariz_Picon_2007,Meerburg2009,Agullo:2010ws,Kundu:2011sg,Ghosh_2023}.
There are efforts in the literature to explore the observational
effects of evolving primordial perturbations from
non-Bunch-Davies
states~\cite{Martin:2000xs,Koh:2004ez,Holman:2007na,Kundu:2011sg,Agarwal:2012mq,Aravind:2013lra,Kundu:2013gha,Ashoorioon:2013eia,Ashoorioon:2014nta,Chen_2014,Dona:2016ohf,Ragavendra:2020vud,Fumagalli:2021mpc,Kanno:2022mkx,Akama:2023jsb,Mondal:2024glo}.
Such models can lead to certain features in the power spectrum and
often enhance the bispectrum in the folded
limit~\cite{Chen_2007,Meerburg2009,Agullo:2010ws,Meerburg:2010rp,Ganc:2012ae,Shiraishi_2013,Chen_2014,Chen_2015,Akama:2020jko,Akama:2023jsb}.
Besides, they also lead to backreaction that could jeopardize
accelerated expansion, unless circumvented
appropriately~\cite{Holman:2007na,Kundu:2013gha,Ragavendra:2020vud,Fumagalli:2021mpc}.
However, such efforts often assume perturbations to have zero
expectation value in their initial state. In this paper, we consider
a state that can induce a non-zero one-point function for
perturbations.

A non-trivial one-point function for the primordial perturbation,
\emph{a primordial mean}, with generic spatial dependence means
violation of statistical homogeneity and hence of ergodicity (see
Appendix~D of~\cite{weinberg2008}). This means both a modification
of N-point functions and their interpretation vis-a-vis cosmological
data (see supplemental material for detailed discussion).

In this work, we examine the observational implications of the
violation of both statistical homogeneity and isotropy. For specific
implementation, we consider the initial quantum state of scalar
perturbations to be a coherent state, instead of the Bunch-Davies
vacuum, during inflation. This state induces a non-zero,
space-dependent one-point function. It violates the translational
invariance at the perturbative level and therefore leads to statistical
inhomogeneity.

We first discuss the form of the primordial mean for scalar
perturbations arising from the coherent initial state. Although we
focus on one-point function induced by a coherent state, the key
arguments and results are applicable to any scenario where the
perturbations acquire a non-zero space-dependent mean.

\noindent
\underline{\it Coherent states for primordial perturbations}--- We
consider slow-roll inflation, where the background dynamics is
captured by the Hubble parameter $H$ and the first slow-roll
parameter
$\epsilon_1 = -\dot H/H^2 \ll
1$~\cite{Liddle:1994dx,Riotto:2002yw,Baumann2009}.\footnote{Overdots
denote $\d/\d \eta$. $H$, $\epsilon_1$ are treated as constants,
albeit they are weakly time-dependent.} In our discussion, we do
not specify the dynamics of the background beyond these parameters,
such as evolution of inflaton field(s) or shape of the driving
potential. We shall focus on the perturbations evolving over such a
background and the properties of their correlation functions.


\textit{One-point function}--- We quantize the gauge-invariant
comoving curvature perturbation $\cR(\eta, \vx)$, using the
annihilation and creation operators, $a_\vk$ and $a_\vk^\dagger$
as~\cite{Stewart:1993bc,Martin:2004um,Sriramkumar:2009kg}
\begin{equation}
\cR(\eta, \vx) = \int \f{\d^3\vk}{(2\pi)^{3/2}}
\big[a_\vk \, f_k(\eta) \, e^{i \vk \cdot \vx} + 
a_{\vk}^\dagger \, f_k^\ast(\eta) \, e^{-i \vk \cdot \vx}\big]\,,
\end{equation}
where $f_k(\eta)$ is the mode function associated with the comoving
wavenumber $k$.\footnote{$\eta\equiv$ conformal time, $\vx \equiv$
  comoving distance coordinate.} The crucial point of this work is
to evolve the perturbations from a coherent state instead of the
Bunch-Davies vacuum. The coherent state is defined as the eigenstate
of the annihilation
operator~\cite{Loudon:2000,Kundu:2011sg,Kundu:2013gha,Mondal:2024glo}
\begin{eqnarray}
a_\vk \vert {\rm C} \rangle &=& {\rm C}(\vk) \vert {\rm C} \rangle\,,
\end{eqnarray}
where $C(\vk)$ is the eigenfunction for a given $\vk$. The creation
and annihilation operators satisfy the standard commutation
relation. Under slow-roll approximation, $f_k(\eta)$
becomes~\cite{Baumann2009,Ragavendra:2020vud,Kundu:2011sg,Kundu:2013gha}
\begin{eqnarray}
f_k(\eta) & \simeq & -\f{H\eta}{\Mpl\sqrt{4k\epsilon_1}}
{\rm e}^{-i k\eta}\left(1 - \f{i}{k \eta} \right)\,,
\end{eqnarray}
where $\Mpl$ is the reduced Planck mass. Note that there is no
mixing of positive and negative frequency components, as it happens
in squeezed states (see supplemental material for discussion
contrasting squeezed and coherent states). We shall focus on the
super-Hubble behavior of the modes, where $f_k$ settles to a
constant and can be related to observables. The
expectation value of $\cR_\vk$ in the coherent state, with the above
mode function in the super-Hubble regime is
\begin{eqnarray}
\langle \cR_\vk \rangle &=& i\l[\f{2\pi^2}{k^3}\ps(k)\r]^{1/2}\alpha(\vk)\,,
\label{eq:rave}
\end{eqnarray} 
where we define
$\alpha(\vk) \equiv {\rm C}(\vk) - {\rm C}^\ast(-\vk)$. The reality
condition on $\braket{\cR(\eta,\vx)}$ implies
$\alpha^\ast(-\vk) = -\alpha(\vk)$. We have also used
\begin{equation}
\ps(k) = \f{H^2}{8\pi^2\Mpl^2\epsilon_1},
\label{eq:ps_sr_bd}
\end{equation}
where $\ps(k)$ is the dimensionless power spectrum
that is obtained in slow-roll inflation, when perturbations are
evolved from Bunch Davies vacuum.\footnote{$H$ and $\epsilon_1$ are
evaluated at Hubble exit of each mode and their time dependences
lead to the minor scale dependence of the spectrum.}

This exercise gives us the mean of curvature perturbation,
$\langle\cR(\vx)\rangle$, that is a non-zero, space-dependent
function. It explicitly violates the assumption of statistical
homogeneity. The scale-dependent function $\alpha(\vk)$ quantifies
the statistical inhomogeneity arising from the coherent state. There
are earlier works that have examined this issue, but with additional
constraints on $\cR$ that preserve statistical
homogeneity~\cite{Kundu:2011sg,Kundu:2013gha,Yusofi:2014vca,Dona:2016ohf,Ghosh_2023}.
We do not assume such additional constraints in our analysis. Such a
mean of a gauge-invariant variable $\langle{\cR(\vx)}\rangle$, can
neither be absorbed in the background nor set to zero using a gauge
transformation. Hence, it is a statistical property of the
perturbation and not an artefact.

The quantity $\alpha(\vk)$ is of dimension $L^{3/2}$. It is a
complex function of the wavevector $\vk$, and hence implies
violation of both statistical homogeneity and isotropy of the
perturbations. If it is a function of the wavenumber alone,
$\alpha(k)$, then it violates statistical homogeneity but preserves
isotropy. Besides, $\braket{\cR(\eta,\vx)}$ vanishes at early times
due to the sub-Hubble behavior of $f_k(\eta)$ and the additional
energy density it gains due to coherent state is insignificant~(see
supplemental material, which contains~\cite{Parker_1974,Bunch_1980,Allen_1987,Abramo_1997,Kundu:2011sg}).


\textit{Two-point correlation}--- We restrict ourselves to the
Gaussian perturbations and do not consider any source of non-Gaussianity in 
$\cR$~\cite{Maldacena:2002vr,Bartolo:2004if,Seery:2005wm,Chen:2006nt,Chen:2010xka,10.1093/ptep/ptu060}.
The two-point correlation of $\cR_\vk$ evaluated in the coherent
state is (see supplemental material)
\begin{eqnarray}
\langle \cR_{\vk_1} \cR_{\vk_2} \rangle &=& \f{2\pi^2}{k_1^{3/2}k_2^{3/2}}
\ps(k_1)\left[\delta^{(3)}(\bm k_1 + \bm k_2)\right. \nn \\ 
& & - \left.\alpha(\vk_1)\alpha(\vk_2) \sqrt{\f{\ps(k_2)}{\ps(k_1)}}\right]\,.
\label{eq:2ptfn}
\end{eqnarray}
The first term, containing $\delta^{(3)}(\vk_1+\vk_2)$, is the
statistically homogeneous and isotropic irreducible part of the
two-point function. The second term containing
$\alpha(\vk_1)\alpha(\vk_2)$ is the reducible contribution due to
the non-zero one-point function, which is explicitly inhomogeneous and
anisotropic. In the extreme case where the first term vanishes, the
two-point correlation is completely given by the product of the
primordial mean. We may call it as the {\it maximally inhomogeneous
case}, where the correlations of perturbations are completely
deterministic.


\textit{Three-point correlation}--- The primordial three-point
function of the scalar perturbations in the presence of a non-zero
mean becomes
\begin{eqnarray}
\langle \cR_{\vk_1} \cR_{\vk_2} \cR_{\vk_3}\rangle &=& 
\langle \cR_{\vk_1} \cR_{\vk_2} \cR_{\vk_3}\rangle_{\rm irr}
\nn \\
& & + \big[\langle \cR_{\vk_1}\rangle \f{2\pi^2}{k_2^3}\ps(k_2)\delta^{(3)}(\vk_2+\vk_3) \nn \\
& & +~\text{two permutations} \big] \nn \\
& & + \langle \cR_{\vk_1}\rangle\langle \cR_{\vk_2}\rangle\langle \cR_{\vk_3}\rangle\,.
\label{eq:3ptfn}
\end{eqnarray}
The first term is the statistically homogeneous and isotropic irreducible 
contribution. It may arise from cubic (or higher order) action of $\cR$~\cite{Martin:2011sn,Arroja:2011yj,Hazra:2012yn,Ragavendra:2020old,Ragavendra:2023ret} 
and we shall ignore it as mentioned earlier. The subsequent terms
arise due to the one-point function. There are three terms correlating 
the irreducible two-point and the one-point functions, while the last 
term is the completely reducible contribution (see supplemental material). 
As we argue in the next section, the current data constrains the 
dimensionless quantity: $k^{3/2} \vert\alpha(\vk)\vert < 1$. 
Therefore, we expect the last term $\propto \alpha^3$ to be subdominant, 
reducing the three-point correlation to
\begin{eqnarray}
\langle \cR_{\vk_1} \cR_{\vk_2} \cR_{\vk_3}\rangle & \simeq & 
\langle \cR_{\vk_1}\rangle \f{2\pi^2}{k_2^3}\ps(k_2)\delta^{(3)}(\vk_2+\vk_3)
\nn \\ & & +~\text{two permutations}\,.
\label{eq:3ptfn2}
\end{eqnarray}

\noindent
\underline{\it Observational consequences}--- Two main observables
that could probe the modifications to two- and three-point functions
are galaxy surveys and CMB anisotropies. Eqs.~(\ref{eq:2ptfn})
and~(\ref{eq:3ptfn2}) show that both the diagonal and off-diagonal
($\vk_1\ne\vk_2$) parts of correlation are modified. The main data
products in both cases are diagonal correlations ($\vk_1 = -\vk_2$
averaged over a spherical shells of fixed $|\vk_1|$ and
$\ell' = \ell$ averaged over all $m$ and $m'$) along with the
covariance matrix. The diagonal terms of the covariance matrix
determine the error on the measured quantities while the
off-diagonal elements represent the impact of various systematics
(e.g.\ incomplete sky coverage, complicated survey geometry,
foregrounds, noise leakage, etc.). This is the case for
statistically homogeneous and isotropic perturbations. The quantity
that is relevant for these observational probes is the dimensionless
power spectrum which is the prefactor of $\delta^{(3)}(\vk_1+\vk_2)$
in Eq.~\eqref{eq:2ptfn}.

To define an equivalent dimensionless quantity that captures the
complete two-point function in our case of inhomogeneous
perturbations, we consider the combination 
${[(k_1k_2)^{3/2}}/{(2\pi^2)}]\langle {\cal R}_{\vk_1} {\cal
  R}_{\vk_2} \rangle$ and integrate it using Eq.~\eqref{eq:2ptfn}
over suitable range of $\vk$. To examine the diagonal terms of the
correlation function $(\vk_2 = -\vk_1)$, we integrate over
$\vk_2$ around a sector of a shell, at a radius $k_1$, in the
direction of $-\hat \vk_1$, with a small volume $k_1^3\Delta^3$. Let
us call this quantity $\sps(\vk,-\vk)$ which simplifies to
\begin{eqnarray}
\sps(\vk,-\vk) & \simeq & \ps(k)\bigg[1 + (k\Delta)^3 \, \vert \alpha(\vk) \vert^2\bigg]\,,
\label{eq:sps_kk}
\end{eqnarray}
for sufficiently small $\Delta$. Notice that $\sps(\vk,-\vk)$ is
guaranteed to be real and positive. The amplitude and scale
dependence of $\alpha(\vk)$ adds to $\ps(k)$ and hence is degenerate
with its behavior.

The contributions unique to $\alpha(\vk)$ are the off-diagonal terms
of the correlation matrix, $\sps(\vk_1,\vk_2)$ with
$\vk_1 \neq -\vk_2$. We know the correlation function is symmetric
as $\sps(\vk_1,\vk_2)=\sps(\vk_2,\vk_1)$. To compute it, we
integrate the two-point function over $\vk_1'$ around $\vk_1$ and
$\vk_2'$ around $\vk_2$ over small volumes of $k_1^3\Delta^3$ and
$k_2^3\Delta^3$ respectively and obtain their average:
\begin{equation}
\sps(\vk_1,\vk_2) \simeq - \f{(k_1^3+k_2^3)}{2}\Delta^3\sqrt{\ps(k_1)\ps(k_2)} \alpha(\vk_1)\alpha(\vk_2)\,.
\label{eq:sps_k1k2}
\end{equation}
This quantity can be complex and is completely determined by the
statistically inhomogeneous component of the perturbations. Having
defined the dimensionless equivalent of the power spectrum for our
case, we first consider implications for galaxy surveys.

\textit{Galaxy surveys}---Galaxy surveys directly measure the matter
power spectrum. The matter power spectrum measure the two-point
correlation of the galaxy density field $\delta_g(\vk)$. The
late-time density field (measured at $t= t_f$; for SDSS galaxy
survey it corresponds to $z\simeq 0.3$) is related to the primordial
curvature perturbation:
\begin{equation}
\delta_g(\vk)  =  \f{2}{5\Omega_m}\left(\f{k}{ H_0}\right)^2 b \cRk T(k,t_f)\,.
\end{equation}
Here $T(k,t_f)$ is the transfer function obtained from the solutions
of the multi-fluid Einstein-Boltzmann equations at $t= t_f$; $b$ is
the galaxy distribution bias vis-a-vis the matter
distribution~\cite{dodelson2021}.

The power spectrum of galaxies can be measured using the unbiased
estimator~\cite{dodelson2021}:
\bea
\hat P_g(k_i) &=& \f{1}{m_i}\sum_{\vk}^{\vert \vert \vk\vert
  -k_i \vert < \Delta k/2} \delta_g(\vk) \delta_g(-\vk) - P_N\,.
\label{eq:galpowest}
\eea
Here $m_i$ is the number of modes in a given bin of range
$k$ to $k + \Delta k$ and the sum is carried out for these modes,
and $\delta_g(\vk)$ is the observed overdensity of the distribution
in Fourier space. $P_N \simeq 1/n$, where $n$ is the number density
of galaxies in the survey volume, is the Poisson noise. We neglect
the impact of Poisson noise in the rest of this section.

The ensemble average of galaxy overdensity follows from Eq.~(\ref{eq:2ptfn}):
\bea
\langle \delta_g(\vk) \delta_g(-\vk) \rangle   = P_g(k)\delta_{\vk,-\vk} + \vert \bar \delta_g(\vk) \vert^2\,,
\label{eq:powave}
\eea
with
\begin{equation}
\bar \delta_g(\vk) = \f{2i\pi}{5\Omega_m}\left(\f{k}{H_0}\right)^2 b T(k,t_f) \sqrt{\f{2\ps(k)}{k^3}}\alpha(\vk)\,.
\end{equation}
Using Eqs.~(\ref{eq:galpowest}) and~(\ref{eq:powave}), the
covariance of the estimator is computed as~\cite{dodelson2021}
\begin{equation}
{\rm Cov}[\hat P_g(k_i)\hat P_g(k_j)] 
= \f{2}{m_i} P^2_g(k_i) \left[1 + 2\f{\vert \bar \delta_g(k_i) \vert^2}{P_g(k_i)}\right] \delta_{ij}\,.
\label{eq:covmat}
\end{equation}
For simplicity, we have assumed $\alpha(\vk)=\alpha(k)$ and so
$\bar \delta_g(\vk)=\bar \delta_g(k)$ in this estimate.
Eq.~(\ref{eq:covmat}) shows that even with off-diagonal two-point
correlations, the covariance matrix is diagonal. But the statistical
inhomogeneity modifies both the measured diagonal correlation
[Eq.~\eqref{eq:powave}] and the estimated covariance matrix
[Eq.~\eqref{eq:covmat}]. Interestingly, the covariance matrix is
proportional to the statistically homogeneous component of the
signal, and therefore its non-zero value already rules out the {\it
  maximally inhomogeneous} case mentioned earlier. If we require the
statistically inhomogeneous component to be sub-dominant, we get an
approximate bound of $k^{3/2} \vert\alpha(k)\vert < 1$
[cf.~Eq.~\eqref{eq:sps_kk}].

However, it is clear that the current data products are not suitable
for revealing the statistically inhomogeneous component of the
signal. The most direct way to isolate this component is to measure
the cross-correlation $P(\vk_1, \vk_2)$ [Eq.~(\ref{eq:sps_k1k2})]
and its covariance matrix.

\textit{CMB anisotropies}--- As in galaxy surveys, the amplitude and
behavior of $\alpha(\vk)$ can also be observationally constrained
using cross-correlation of modes in the CMB angular power spectrum,
$C_{\ell,\ell'}$. The correlations in CMB anisotropies are modified
in our case:
$\Braket{a_{\ell m} a_{\ell' m'}} = C_\ell \delta_{\ell\ell'}
\delta_{mm'} + \Braket{a_{\ell m}} \Braket{a_{\ell' m'}}$, where
\begin{equation}
\Braket{a_{\ell m}} = 2^{5/2} \pi^2 \, i^{\ell+1}
\int \frac{\d^3 \vk}{(2 \pi)^{3/2}} \Theta_\ell(k) \sqrt{\ps(k)} \frac{\alpha(\vk)}{k^{3/2}} Y^*_{\ell m}(\hat{\vk})\,.
\label{eq:alm}
\end{equation}
For $\alpha(\vk)=\alpha(k)$, i.e.\ for statistically inhomogeneous
but isotropic perturbations, $\Braket{a_{\ell m}}=0$. Therefore, CMB
anisotropies are insensitive to statistical inhomogeneities, unless
they are anisotropic.

Both WMAP and Planck data have been analyzed to detect the mean of
temperature fluctuations, $\Braket{a_{\ell m}}$. There is weak
evidence of non-zero mean over 
$61\leq \ell \leq 86$~\cite{Armendariz_Picon_2011,Ichiki:2014rua}
and $213\leq \ell \leq 256$~\cite{Ichiki:2014rua,Kashino:2011ih}.
Given the sensitivity of such a measurement to unknown systematics,
it is hard to connect such results to theory. However, based on
these analyses, it is safe to assume that the overall mean is small and
within the standard deviation of the CMB temperature
fluctuations~\cite{Armendariz_Picon_2011}.

If we ignore the off-diagonal terms and focus on the diagonal part
of $\sps(\vk_1,\vk_2)$ or $C_{\ell,\ell'}$ we may utilize the
existing constraints on the primordial scalar power spectrum
$\ps(k)$ from various cosmological datasets to broadly constrain
$\sps(\vk,-\vk)$. We perform such a comparison by using specific
functional forms of $\alpha(\vk)$ in the appendix.

Besides, the CMB anisotropies have been analysed to constrain
anisotropic component of the primordial power
spectra~\cite{Prunet:2004zy, Souradeep_2006, Gordon_2007,
  Carroll:2008br, Bartolo_2012, Rath_2015, Kothari_2016}. The
statistical anisotropy is typically investigated by introducing a
special direction ($\hat{\bm{d}}$) in $\ps(k)$ as
$\ps(\bm{k}) = \ps^0(k)[1 + g_\ast (\hat{\bm{k}} \cdot
\hat{\bm{d}})^2]$, where $\ps^0(k)$ is the standard power spectrum
and $g_\ast$ captures the anisotropic
contribution~\cite{Ackerman:2007nb, Schwarz_2016}. From the Planck
data, we have $g_\ast = 0.002 \pm 0.016$ at $68\%$
C.L.~\cite{Kim:2013gka,Planck:2015sxf,Planck:2018jri,Schwarz_2016,Bartolo:2015dga}.
Comparing such $\ps(k)$ with $\sps(\vk,-\vk)$ in
Eq.~\eqref{eq:sps_kk}, one can constrain the magnitude of
$\alpha(\vk)$. In our case, the role of $g_\ast$ is played by
$k^3\Delta^3\vert\alpha(\vk)\vert^2$ [cf.~Eq.~\eqref{eq:sps_kk}]. If
we set $\Delta=0.1$, then $\vert g_\ast\vert \leq 10^{-2}$, implies
\bea
k^{3/2}\vert\alpha(\vk)\vert \leq \sqrt{10}\,.
\label{eq:cmbconan}
\eea

Notice that, if we choose parameters such that
$k^{3/2}\alpha(\bm k)$ is nearly a constant, the bound on
$k^{3/2}\alpha(\bm k)$ is rather weak as the overall normalization
cannot be determined with the same level of
precision~\cite{Planck:2019evm}.

\noindent
\underline{\it Implications for spectral distortion}--- At smaller
scales, $\alpha(\vk)$ has interesting implication for CMB spectral
distortion. If energy is injected into CMB in the redshift range
$10^3 < z < 10^6$, it causes spectral distortion of CMB. In this
redshift range, sound waves of the coupled baryon-photon system are
damped by Silk damping for $0.1 < k/\mpcinv < 10^4$. Here we are
interested in the early release of energy, which causes $\mu$
distortion, as this range of scales is not constrained by other
cosmological observables. As the amplitude of the sound wave is
determined by primordial matter power spectrum, an enhancement is
constrained by current FIRAS bounds on CMB distortion.

The CMB distortion parameter $\mu(\bm x)$ can be related to
primordial perturbations as~\cite{Chluba:2015bqa}
\begin{eqnarray}
\mu(\bm x) & = & \int \int \f{\d^3 \vk_1 \d^3 \vk_2}{(2\pi)^3}\, \cR_{\vk_1} \cR_{\vk_2} 
{\rm e}^{i (\vk_1 + \vk_2) \cdot \bm x}\, \nn \\
& & \times W_\mu(k_1)W_\mu(k_2)\,,
\label{eq:mudef}
\end{eqnarray}
where $W_{\mu}(k)$ is the window function that captures the range of
scales impacted by the acoustic damping of primordial perturbations
(see~\cite{Chluba:2015bqa} for the exact form). Using
Eq.~(\ref{eq:2ptfn}), the ensemble average of $\mu(\bm x)$
[Eq.~\eqref{eq:mudef}] is
\begin{eqnarray}
\Braket{\mu(\bm x)} & \simeq & \int \d \ln k\, \ps(k) \, W^2_\mu(k) \nn \\
& & + \left[ \int \f{\d^3 \vk}{(2\pi)^{3/2}} \f{\alpha(\vk)}{k^{3/2}} 
\sqrt{\ps(k)}
{\rm e}^{i \vk \cdot \bm x}\, W_\mu(k) \right]^2.
\label{eq:muave}
\end{eqnarray}
The space-independent first term is the same as in the usual case,
while the additional space-dependent contribution is due to statistical inhomogeneity.

The additional signal depends on the magnitude of $\alpha(\vk)$ and
its angular properties. The observed signal arises from a redshift
range $10^5 <z < 10^6$ and is a function of angle on the sky. The
plane-wave expansion in Eq.~(\ref{eq:mudef}) and subsequent spatial
averaging can be used to isolate the statistically anisotropic
component. The statistically isotropic signal only contributes to
the global signal as it corresponds to $\vk_1 = -\vk_2$. The
anisotropic part of the signal would be proportional to
$\int \d \Omega_{\hat{\bm{k}}} \alpha(\vk) Y^\ast_{\ell m}(\hat \vk)$
[similar to Eq.~\eqref{eq:alm}]. For $\alpha(\vk)=\alpha(k)$, i.e.\
for statistically inhomogeneous but isotropic mean, the only
contribution is from $\ell = 0$ mode, which corresponds to enhancing
the global signal. Using specific parameterizations in appendix, 
Figs.~\ref{fig:ps_ln_contours} and~\ref{fig:ps_pl_contours} capture
the impact of this component at small scales. The best bounds of
$\mu$ have been obtained using COBE-FIRAS, which had a beam of
nearly $10^\circ$~\cite{Fixsen:1996nj}. Future probes such as PIXIE
will improve on both sensitivity and angular
resolution~\cite{Kogut:2011xw}.
 
\noindent
\underline{\it Bispectrum}--- The impact of primordial mean on bispectrum 
can be computed from both the galaxy and CMB data. 
The CMB data probes large scales and hence is directly relevant for
probing primordial physics. As noted in the discussion on two-point
functions, the data products in this case also focus on obtaining
the statistically homogeneous-isotropic component of the signal. In
particular, the triangular configurations
$\vk_1+\vk_2 + \vk_3 = {\bm 0}$ and covariance matrix based on these
triangles are used to obtain the bispectrum and its variance. One of
the limits often considered in data analysis is the squeezed
limit, with $\vk_1 \to \bm 0$ and $\vk_2 \simeq - \vk_3$. We argue
that this limit is comparable to the form of bispectrum obtained in
Eq.~\eqref{eq:3ptfn2}, and therefore, we consider only this limit
for comparing our theoretical predictions with data. In this limit,
only the first term of Eq.~\eqref{eq:3ptfn2} with
$\delta^{(3)}(\vk_2+\vk_3)$ contributes, whereas the other terms
containing
$\delta^{(3)}(\vk_1+\vk_{2,3}) \simeq \delta^{(3)}(\vk_{2,3})$ do
not contribute at finite values of $k_2$ or $k_3$. This gives us
\begin{eqnarray}
\langle \cR_{\vk_1 \to 0} \cR_{\vk_2} \cR_{\vk_3}\rangle & \simeq & 
\f{2\pi^2}{k_2^3} \langle \cR_{\vk_1} \rangle \ps(k_2) \delta^{(3)}(\vk_2+\vk_3)\,. \nn \\
\end{eqnarray}
This behavior of three-point function mimics that of the squeezed
limit of the bispectrum arising in a statistically homogeneous case,
\begin{eqnarray}
\langle \cR_{\vk_1 \to 0} \cR_{\vk_2} \cR_{\vk_3}\rangle = (2\pi)^3
{\cal B}_{\rm sq}(\vk_1,\vk_2)\delta^{(3)}(\vk_2+\vk_3)\,.
\end{eqnarray}
Comparing the two expressions, we write
${\cal B}_{\rm sq}(\vk_1,\vk) = \langle \cR_{\vk_1} \rangle
{\ps(k)}/{(4\pi k^3)}$ for our case, although a subtle difference is
that $k_1$ is not taken to be zero but the smallest wavenumber of
interest $10^{-4}\,\mpcinv$ for observational purposes. We emphasize
that the distribution of $\cR(\vx)$ is still Gaussian and the
computed bispectrum is just the contribution to the three-point
function from the non-zero mean.

Using the bispectrum, we can compute the corresponding
non-Gaussianity parameter $\fnl$~\cite{Maldacena:2002vr}:
\begin{eqnarray}
\fnl^{\rm sq} &=& -\f{5}{3}\sqrt{2\pi}\f{k_1^3k^3 {\cal B}_{\rm sq}(\vk_1,\vk)}{\ps(k_1)\ps(k)}\,, \\
& \simeq & -\f{5\,i}{6}\sqrt{\f{\pi}{\ps(k_1)}}
k_1^{3/2}\alpha(\vk_1)\,.
\label{eq:fnlexp}
\end{eqnarray}
Notice that $\fnl \propto \alpha(\vk_1)$ and hence is a direct probe
of the nature of the primordial mean. Contrast this against the
$\fnl$ predicted by the consistency relation
$\fnl^{\rm sq} = (5/12)(\d \ln \ps(k)/\d \ln k)$ which is typically
small and scale independent~\cite{Maldacena:2002vr}. Clearly, the
behavior we obtain deviates from this prediction. The $\fnl$ we
obtain is a complex quantity and if we consider only its magnitude,
we can compare it against the existing bound on the local template
of $\fnl$ from the Planck data: $\fnl^{\rm loc}=-0.9 \pm 5.1$ at
$68\%$ C.L.~\cite{Planck:2019kim}. Substituting the value of
spectrum at $k_1 = 10^{-4}\,\mpcinv$, i.e.
$\ps(k_1) \simeq 2.7\times 10^{-9}$, we obtain an upper bound on
amplitude of $\alpha(\vk)$ over the largest of CMB scales:
\begin{eqnarray}
k_1^{3/2}\vert \alpha(\vk_1)\vert & \leq & 1.8 \times 10^{-4}\,.
\label{eq:fnlcons}
\end{eqnarray}
This is the strongest constraint on $\alpha(\vk)$ from large scale
observables. Note that the scale dependence of $\fnl$ arises solely
from the squeezed wavenumber $k_1$.

There are also weak constraints from Planck data on the anisotropic
nature of $\fnl$ for certain templates motivated by parity-violating 
models of early universe~\cite{Planck:2015zfm,Planck:2019kim}. As the 
bispectrum is anisotropic in our case, this is yet another probe of our 
proposed scenario.

\noindent
\underline{\it Conclusions}--- We investigate the observable consequences of 
breakdown of statistical homogeneity and isotropy of cosmological perturbations. 
We illustrate it by having coherent state as the initial quantum state
for primordial perturbations.\footnote{It is also possible to break
statistical homogeneity and isotropy in any initial quantum state
that is mix of ground and excited states.} We construct a
non-zero, space-dependent primordial mean $\alpha(\vk)$, and compute
the corresponding two- and three-point functions. Our main results
are summarized below:
\begin{enumerate}
\item The two- and three-point correlations of observables acquire
  reducible contributions. While the galaxy distribution is
  sensitive to both statistical inhomogeneity and anisotropy, CMB is
  specifically sensitive to the statistically anisotropic component
  of the mean.
\item The off-diagonal components in the two-point correlation of
  perturbations are uniquely sensitive to, and hence clean probes
  of, the primordial mean.
\item A novel aspect of having a primordial mean is the
  non-vanishing three-point correlation, even for a Gaussian field.
  We show that the bound on $\fnl$ yields the strongest constraint
  on $\alpha(\vk)$~[Eq.~\eqref{eq:fnlcons}].
\item The primordial mean can lead to excess fluctuations over small
  scales while satisfying constraints on large scales (see appendix
  with Figs.~\ref{fig:ps_ln_contours} and~\ref{fig:ps_pl_contours}).
  The acoustic damping of excess perturbations over such scales
  shall increase CMB $\mu$-distortion and render it highly
  anisotropic.
\end{enumerate}

Other implications of enhanced power at small scales are increased
rates of formation of haloes and compact objects such as primordial
black holes (similar to inflationary models with an ultra slow-roll
phase~\cite{Garcia-Bellido:2017mdw,Germani:2017bcs,Atal:2019erb,Motohashi:2019rhu,Ragavendra:2020sop,Ragavendra:2023ret,Ragavendra:2024yfp}).
Evolving primordial tensor perturbations from a coherent state may
generate increased amplitude of stochastic gravitational waves,
which can be constrained by CMB B-modes over large scales, and
probes such as PTA over small
scales~\cite{BICEP:2021xfz,NANOGrav:2023hvm}. 

\noindent
\underline{\it Acknowledgements}--
HVR thanks Raman Research Institute for support through for postdoctoral fellowship where a major part of this work was carried out.
HVR acknowledges support in part by the MUR PRIN2022 Project ``BROWSEPOL: Beyond standaRd mOdel With coSmic microwavE background POLarization"-2022EJNZ53 financed by the European Union - Next Generation EU. 
DM thanks Raman Research Institute for financial support and hospitality through the Visiting Student Program.
DM thanks Kinjalk Lochan for helpful discussions and comments.

\section*{appendix}
\underline{\textit{Parameterization of  primordial mean}}---
The dimensionless combination that arises in the one-point function of 
$\langle \cRk \rangle$ and  in higher-order correlations is
$k^{3/2}\alpha(\bm k)$\, [Eq.~\eqref{eq:sps_kk}]. 
Here we consider parametric forms for this quantity, with the condition that the one-point function vanishes at large scales, $\langle\Ro(\vk)\rangle \to 0$ for $k \to 0$. 

As noted before, both the CMB and galaxy data are sensitive to the diagonal part of the correlation [Eq.~\eqref{eq:sps_kk}]. 
The Planck data is consistent with a nearly scale-invariant
primordial power spectrum over $k\simeq 5\times 10^{-4} \hbox{--}0.2 \, \rm Mpc^{-1}$~\cite{Planck:2018jri}. 
The impact of non-zero scale-dependent $k^{3/2}\alpha(\bm k)$ 
may lead to features in the spectrum which are not strongly favoured by the data.
The near-scale invariant matter power spectrum is also preferred at smaller scales probed by Lyman-$\alpha$ data (e.g. see \cite{Ragavendra:2024yfp} and references therein).
Therefore, it would be safe to assume that $k^{3/2}\alpha(\bm k)$ is constrained to be nearly a constant in range of scales $k\simeq 5\times 10^{-4} \hbox{--}10 \, \rm Mpc^{-1}$. This constraint motivates the 
choice of functional forms of  $k^{3/2}\alpha(\bm k)$, which we consider next. 


\textit{Lognormal form}---
Let us first consider a parameterization that is bounded on both large and small scales. The lognormal form for  $\alpha(\bm k)$ satisfies this requirement:
\begin{eqnarray}
k^{3/2}\alpha(\bm k)&=& -i\,\alpha_0 
\exp\left[-\f{1}{2 \Delta^2_{\rm f}}\ln^2\left(\f{k}{k_0}\right)\right]
\beta(\hat\vk)\,.
\label{eq:lognormal}
\end{eqnarray}
Here, $\alpha_0$ is the magnitude of the statistical inhomogeneity, $k_0$ sets the scale around which inhomogeneity starts playing a role and $\Delta_{\rm f}$ determines the range of scales about $k_0$ where it is relevant.
The function $\beta(\hat \vk \cdot \hat \vk_0)$ captures the violation of statistical isotropy, with a special direction along $\vk_0$. 
It has the property of $\beta^\ast(-\hat \vk) = \beta(\hat \vk)$ and is normalized such that $\vert \beta(\hat \vk) \vert^2 = 1$.
In the absence of any special direction, $\beta(\hat \vk)=1$ and $\alpha(\vk)=\alpha(k)$.
In such a case, $\alpha(k)$ leaves no imprint on the angular power spectrum $C_\ell$ of CMB [cf.~Eq.~\eqref{eq:alm}].

\begin{figure}[]
\includegraphics[scale=0.34]{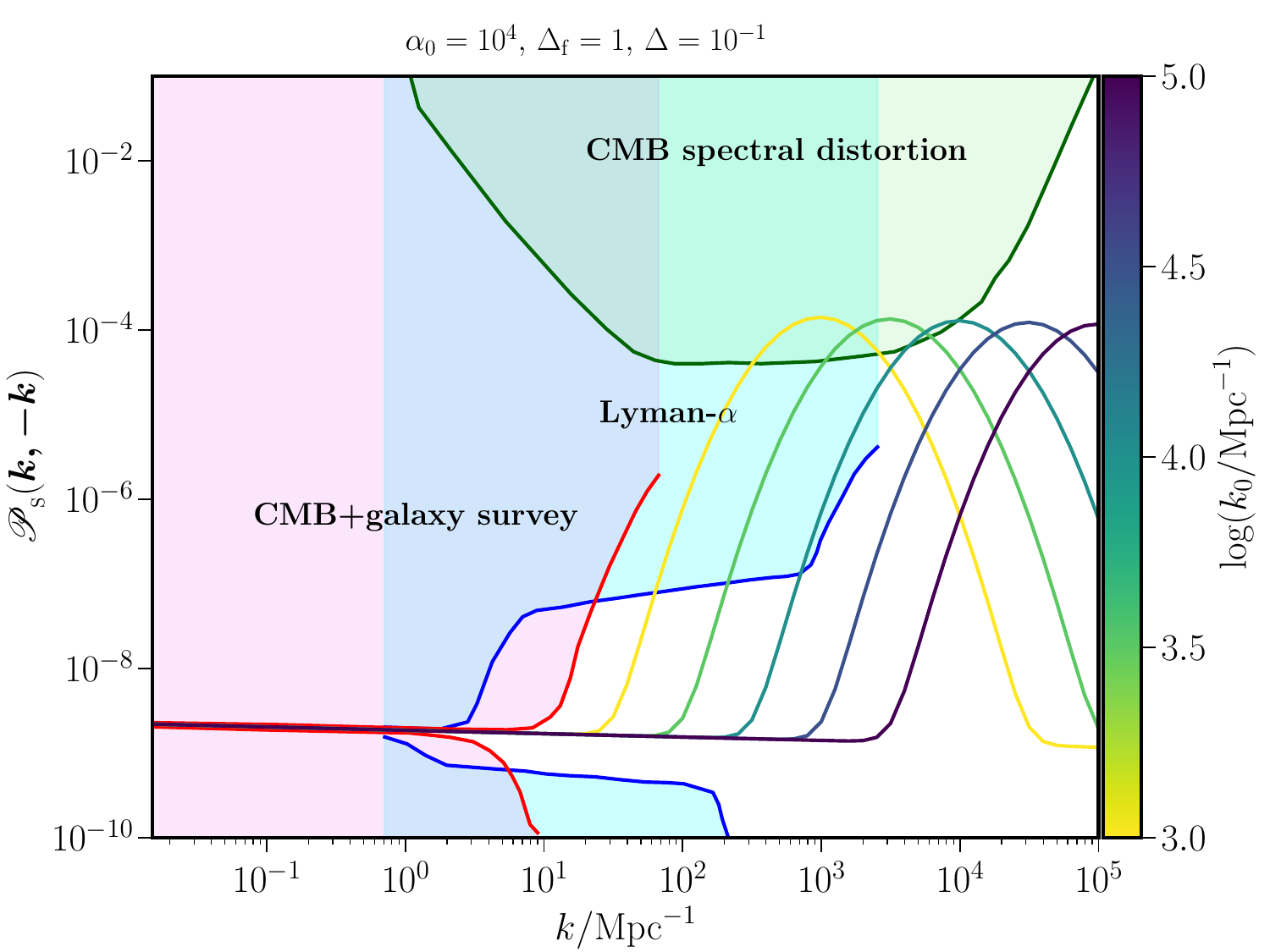}
\includegraphics[scale=0.34]{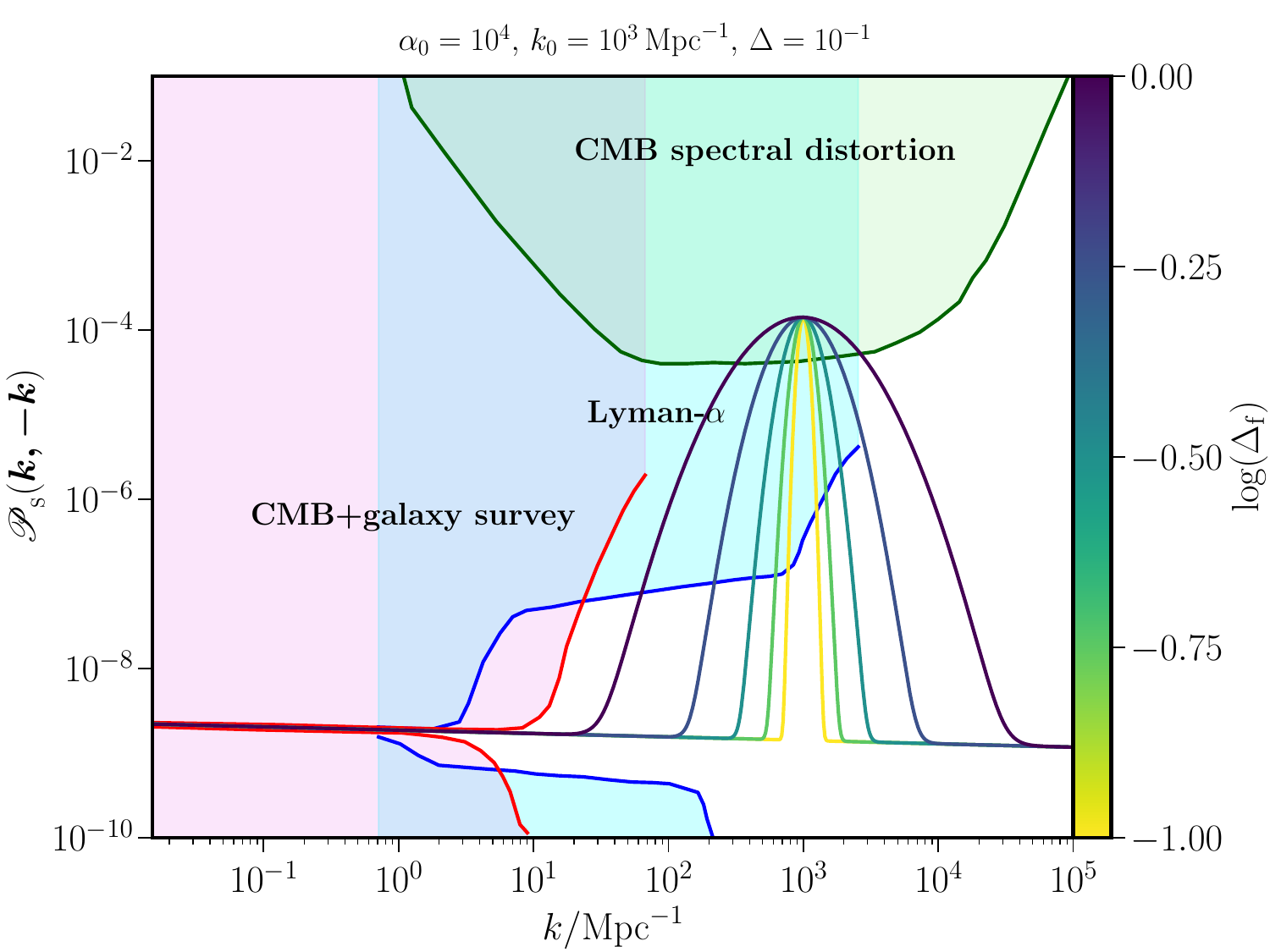}
\vskip -0.1in
\caption{The dimensionless two-point auto-correlation of modes $\sps(\vk,-\vk)$ 
is presented for the lognormal parameterization of $\alpha(\vk)$ across a range of $k_0$ and $\Delta_{\rm f}$ values (on top and bottom panels respectively). 
The values chosen for other parameters are given on top of respective panels.
The constraints presented are taken from the results of Ref.~\cite{Ragavendra:2024yfp}. 
Although the quantity constrained in~\cite{Ragavendra:2024yfp} is $\ps(k)$ and 
there is a minor dependence of contours of constraints on the model used in the
reference, we can obtain reliable upper bounds on the parameters $k_0$ and $\Delta_{\rm f}$ from this illustration.
We may rule out parameter values that lead to deviation over the region constrained by the CMB+galaxy surveys.
However, as $\sps(\vk,-\vk)$ grows over small scales, the corresponding constraints, such as Lyman-$\alpha$, have to be recomputed while accounting for off-diagonal contributions $\sps(\vk_1,\vk_2)$ that become comparable to $\sps(\vk,-\vk)$. 
Hence the existing constraints over those scales shall grow stronger for our model.}
\label{fig:ps_ln_contours}
\end{figure}

With this $\alpha(\vk)$, we compute the complete structure of $\sps(\vk_1,\vk_2)$. 
The diagonal part $\sps(\vk,-\vk)$ [cf.~Eq.~\eqref{eq:sps_kk}] becomes
\begin{eqnarray}
\sps(\vk,-\vk) & = & \ps(k)\Bigg\{1 + \Delta^3 \alpha_0^2
\exp\left[-\f{1}{\Delta^2_{\rm f}}\ln^2\left(\f{k}{k_0}\right)\right]
\Bigg\}\,, \nn \\
\label{eq:sps_kk_ln}
\end{eqnarray}
and the off-diagonal part $\sps(\vk_1,\vk_2)$ [cf.~Eq.~\eqref{eq:sps_k1k2}] is
\begin{eqnarray}
\sps(\vk_1,\vk_2) & = & \f{\Delta^3}{2} \f{(k_1^3+k_2^3)}{(k_1k_2)^{3/2}}\sqrt{\ps(k_1)\ps(k_2)} \beta(\hat \vk_1)\beta(\hat \vk_2) \nn \\
& & \times \alpha_0^2
\exp\left[-\f{\ln^2\left(\f{k_1}{k_0}\right)+\ln^2\left(\f{k_2}{k_0}\right)}{2\Delta^2_{\rm f}}\right]\,.
\label{eq:sps_k1k2_ln}
\end{eqnarray}
We plot the diagonal part $\sps(\vk,-\vk)$ in Fig.~\ref{fig:ps_ln_contours},
in lieu of $\ps(k)$, along with the constraints from cosmological observables such as CMB anisotropies, galaxy surveys, Lyman-$\alpha$ emissions and spectral distortions in CMB. 

\begin{figure}[]
\includegraphics[scale=0.34]{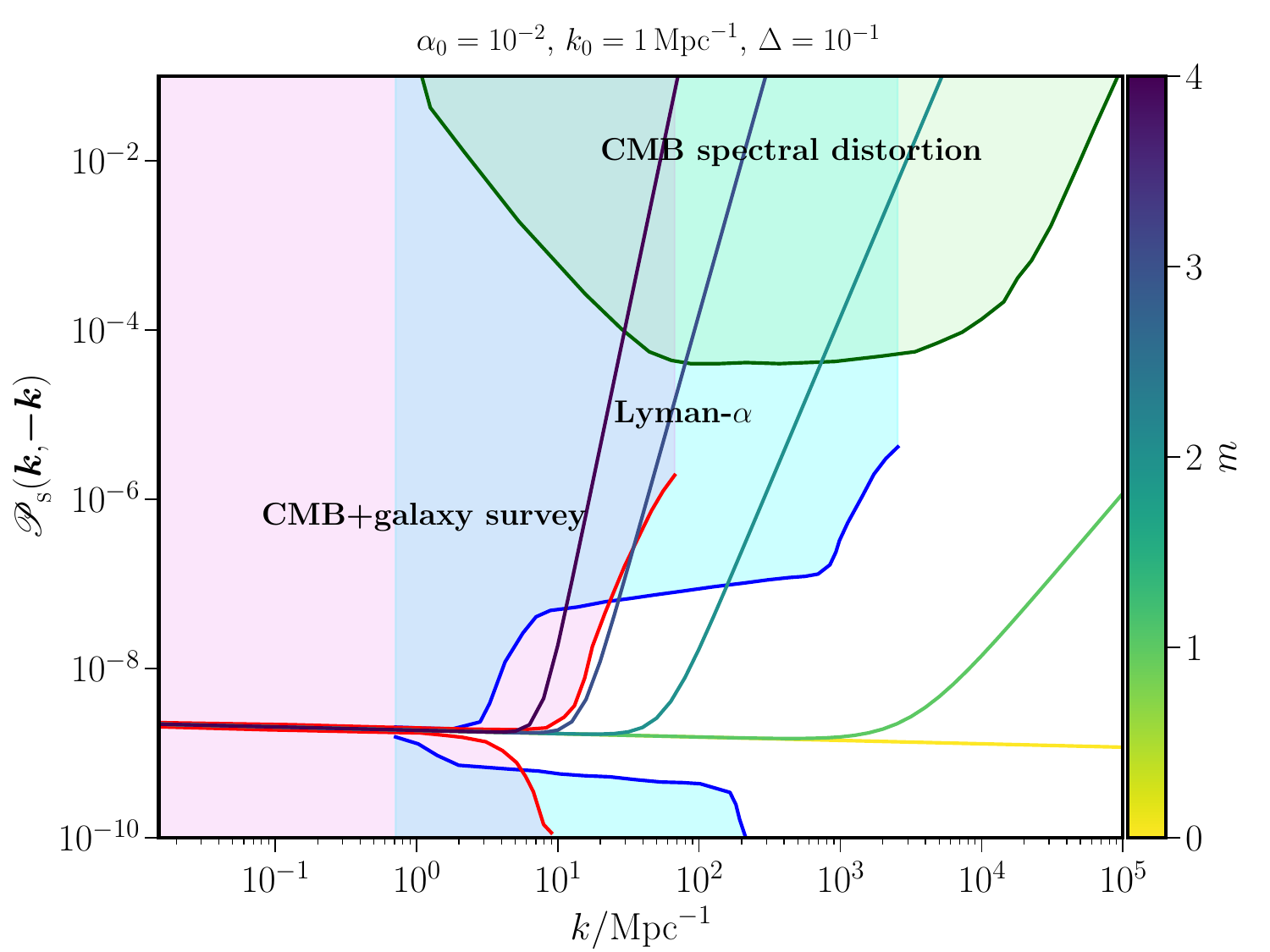}
\includegraphics[scale=0.34]{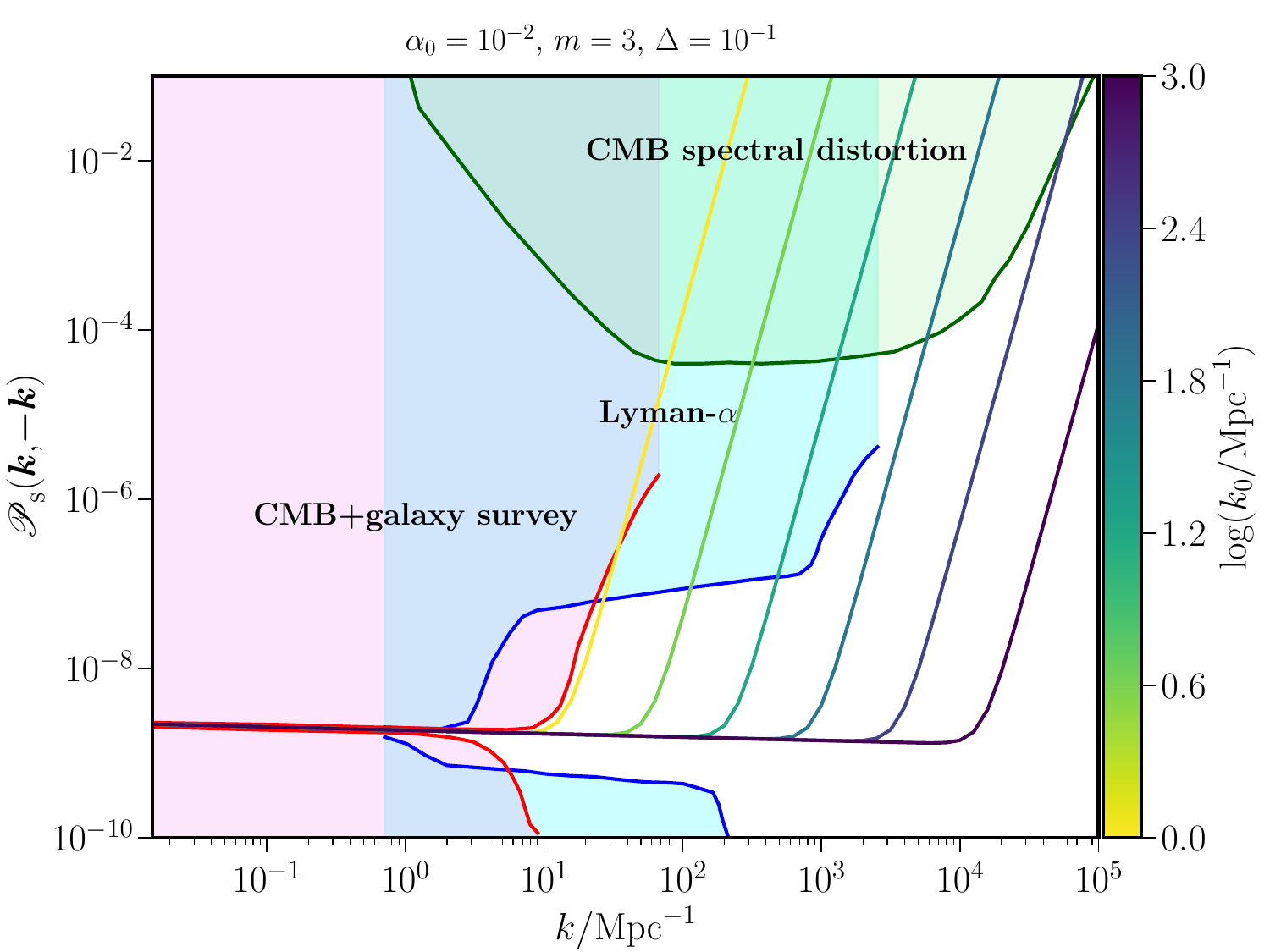}
\vskip -0.1in
\caption{The dimensionless two-point auto-correlation of modes $\sps(\vk,-\vk)$ 
is presented for the power-law parameterization of $\alpha(\vk)$ across a range of $m$ and $k_0$ values (on top and bottom panels respectively). 
The values chosen for other parameters while arriving at respective plots are given on top of each panel.
We have plotted it against the constraint contours as in the previous figure. 
We may safely rule out parameters that deviate over scales constrained by CMB+galaxy surveys, while constraints over small scales have to be recomputed accounting for off-diagonal contribution $\sps(\vk_1,\vk_2)$ that become comparable to $\sps(\vk,-\vk)$.}
\label{fig:ps_pl_contours}
\end{figure}

\textit{Power-law form}---
Another but simpler choice of parameterization of $\alpha(\vk)$ would be a power-law of the form
\begin{eqnarray}
k^{3/2}\alpha(\bm k)&=& -i\,\alpha_0 \left( \f{k}{k_0} \right)^m\beta(\hat{\bm k})\,.
\label{eq:powerlaw}
\end{eqnarray}
The disadvantage of this form is the need for a cutoff scale to prevent the growth of $\alpha(\vk)$ in $k$ from violating the perturbativity of $\braket{\cR(\vx)}$. 
The structure of $\sps(\vk_1,\vk_2)$ with $\vk_1=-\vk_2$ is
\begin{eqnarray}
\sps(\vk,-\vk) & = & \ps(k)\Bigg[1 + \Delta^3 \, \alpha_0^2
\left(\f{k}{k_0} \right)^{2m}\Bigg]\,,
\label{eq:sps_kk_pl}
\end{eqnarray}
and with $\vk_1 \neq -\vk_2$, it is 
\begin{eqnarray}
\sps(\vk_1,\vk_2) & = & \f{\Delta^3}{2} \f{(k_1^3+k_2^3)}{(k_1k_2)^{3/2}}\sqrt{\ps(k_1)\ps(k_2)} \nn \\
& & \times \alpha_0^2 \left(\f{k_1k_2}{k_0^2}\right)^{m}\beta(\hat \vk_1)\beta(\hat \vk_2)\,.
\label{eq:sps_k1k2_pl}
\end{eqnarray}
We illustrate $\sps(\vk,-\vk)$ along with constraints in Fig.~\ref{fig:ps_pl_contours}, as done for the previous parameterization.
\begin{widetext}
\begin{center}
\Large \bf Supplemental material for \\
`Cosmological consequences of statistical inhomogeneity'
\end{center}
\end{widetext}
\section{Primordial correlations in coherent state}
\label{sec:SM1}

In this section, we outline the derivations of the primordial two-
and three-point functions in the coherent state, that are presented in the main text. 
The quantized version of the gauge-invariant curvature perturbation $\cR(\eta,\vx)$ is~\cite{Stewart:1993bc,Martin:2004um,Sriramkumar:2009kg,Baumann2009}
\begin{eqnarray}
\cR(\eta, \vx) 
&=& \int \f{\d^3\vk}{(2\pi)^{3/2}}\cR_\vk(\eta)\, e^{i \vk \cdot \vx}\,, \\
&=& \int \f{\d^3\vk}{(2\pi)^{3/2}}
\big[a_\vk \, f_k(\eta) \, e^{i \vk \cdot \vx} + 
a_{\vk}^\dagger \, f_k^\ast(\eta) \, e^{-i \vk \cdot \vx}\big]\,, 
\nn \\ \\
&=& \f{1}{z}\int \f{\d^3\vk}{(2\pi)^{3/2}}
\big[a_\vk \, v_k(\eta) \, e^{i \vk \cdot \vx} + 
a_{\vk}^\dagger \, v_k^\ast(\eta) \, e^{-i \vk \cdot \vx}\big]\,, \nn \\
\end{eqnarray}
where $z=a\Mpl\sqrt{2\epsilon_1}$ and $v_k$ is the Mukhanov-Sasaki variable related to the mode function as $v_k = f_k\,z$~\cite{Sriramkumar:2009kg,Baumann2009}.
The standard commutation relation governing the creation and annihilation operators is
\begin{eqnarray}
[a_{\vk_1}, a^\dagger_{-\vk_2}] &=& \delta^{(3)}(\vk_1+\vk_2)\,.
\end{eqnarray}
The mode function $f_{k}(\eta)$ obeys the equation of motion
\begin{eqnarray}
f''_k + 2 \f{z'}{z}f'_k + k^2f_k &=& 0\,,
\end{eqnarray}
where the primes denote derivative with respect to $\eta$. Equivalently the Mukhanov-Sasaki equation is
\begin{eqnarray}
v''_k + \left(k^2 - \f{z''}{z}\right) v_k &=& 0\,.
\end{eqnarray}
Under the slow-roll approximation, where $\epsilon_1$ and other higher order slow-roll parameters are much smaller than unity and so $a \simeq -(H\eta)^{-1}$, the general solution for $v_k(\eta)$ is given by
\begin{align}
v_k(\eta) \simeq  \frac{1}{\sqrt{2k}}
\left[ A_{k} \e^{-i k \eta} \left( 1 - \frac{i}{k \eta} \right)
  + B_{k}\e^{i k \eta} \left( 1 + \frac{i}{k \eta} \right) \right],
\label{eq:g_mode_fun}
\end{align}
where $A_{k}$ and $B_{k}$ are two complex functions for a given $k$.
In the standard scenario, the expectation values of the primordial fluctuations are generally evaluated in a `vacuum state'. 
In order to unambiguously specify the vacuum state, one invokes the Bunch-Davies initial condition by setting $A_k=1, B_k=0$. 
In this case, the vacuum state ($a_\vk \ket{0} = 0$) is identified as the minimum-energy eigenstate of the Hamiltonian at the onset of inflation ($\eta \to -\infty$). 

A squeezing operation on this state may lead to non-zero form of $B_k$, thereby also deviating from the minimum-energy criterion.
In our work, we choose the initial state to be the coherent state $\ket{C}$, which is defined as $a_\vk\ket{C}=C(\vk)\ket{C}$.
It just displaces the operators $a_\vk$ and $a^\dagger_\vk$ but does not alter the condition of $A_k=1$ and $B_k=0$ [refer discussion in the next section and references therein].
Thus the solution for $v_k(\eta)$ simply is
\begin{align}
v_k(\eta) \simeq  \f{1}{\sqrt{2k}}
\e^{-i k \eta} \left( 1 - \frac{i}{k \eta} \right)\,,
\end{align}
and the mode function $f_k(\eta)$ becomes
\begin{align}
f_k(\eta) \simeq  -\frac{H\eta}{\sqrt{4k\epsilon_1}}
\e^{-i k \eta} \left( 1 - \frac{i}{k \eta} \right)\,.
\end{align}
Therefore, the one-point function of $\cRk$ evaluated in the super-Hubble limit
($\abs{k \eta} \to 0$) turns out to be
\begin{eqnarray}
\Bra{C} \cRk(\eta) \ket{C}
&\simeq &  \f{i \sqrt{2 \pi^2 \ps(k)}}{k^{3/2}}
\left[ C(\vk) - C^\ast(-\vk) \right],
\end{eqnarray}
where we can identify $\alpha(\vk)=C(\vk)-C^\ast(-\vk)$ and $\ps(k)$ as defined in the main text.
Note that the one-point function in general depends on time. 
Therefore, even if we impose statistical homogeneity and isotropy in the super-Hubble regime by setting $\alpha(\vk)=0$ (for instance, as done in Ref.~\cite{Kundu:2011sg}), $\braket{\cRk}$ may be non-zero in the sub-Hubble regime.

\paragraph{One-point function in real space:}
We shall briefly examine the structure of the one-point function arising from coherent state in real space.
We have obtained that 
\begin{eqnarray}
\braket{\cRk(\eta)} &=& i\left[\f{2\pi^2\ps(k)}{k^3}\right]^{1/2}\alpha(\vk)\,,
\label{eq:one_p_fun}
\end{eqnarray}
in the super-Hubble regime. Let us express
\begin{eqnarray}
\alpha(\vk) = -i\bar\alpha(k)\beta(\hat \vk)\,,
\end{eqnarray}
where we have decomposed $\alpha(\vk)$ as components dependent on amplitude and direction of $\vk$ respectively.
Note that $\alpha^\ast(\vk)=-\alpha(-\vk) \Rightarrow \bar\alpha^\ast(k)\beta^\ast(\hat \vk) = \bar\alpha(k)\beta(-\hat \vk)$ . 
If we assume $\bar\alpha(k)$ is real, then $\beta^\ast(\hat \vk)=\beta(-\hat \vk)$.
If the one-point function is statistically inhomogeneous but isotropic, then $\beta(\hat \vk)=1$.

Using such a behavior of the Fourier mode, we compute the one-point function of the curvature perturbation in real space at sufficiently late times during inflation as
\begin{eqnarray}
\braket{\cR(\vx)} &=& \f{1}{2\sqrt{\pi}} \int \d \ln k \, k^{3/2} \sqrt{\ps(k)} \bar\alpha(k) \nn \\
& & \times \int \d^2 \Omega_\vk \beta(\hat \vk)e^{i\vk\cdot\vx}\,.
\end{eqnarray}
We can perform the angular integral by aligning the $\vx \parallel \hat \vk_z$ and expanding $\beta(\vk)$ in Legendre polynomials. 
The functional form of the resultant integral will depend on the specific directional dependence of $\beta$, say $\theta_0$.

In the simple case of $\beta(\hat \vk)=1$, we can readily show that
\begin{eqnarray}
\braket{\cR(\vx)} &=& 2\sqrt{\pi} \int \d \ln k \, k^{3/2} \sqrt{\ps(k)} \bar\alpha(k) \f{\sin (kx)}{kx}\,.
\end{eqnarray}
Here the $\sin(kx)/(kx)$ serves as a window function with the scale $x$ curtailing the range of $k$ contributing to the integral.
In other cases with non-trivial angular dependence of $\beta(\hat \vk)$, we can express the one-point function as
\begin{eqnarray}
\braket{\cR(\vx)} &=& 2\sqrt{\pi} \int \d \ln k \, k^{3/2} \sqrt{\ps(k)} \bar\alpha(k) T(\theta_0) W(kx)\,, \nn \\
\end{eqnarray}
where $W(kx)$ shall be a window function with behavior similar to $\sin(kx)/(kx)$ and $T(\theta_0)$ captures the special direction that explicitly breaks the isotropy.

On the other hand, it is also interesting to note that the one-point function is insignificant at early times.
At sufficiently early times ($\eta \to -\infty$), using the sub-Hubble behavior of $f_k(\eta)$ we can write $\braket{\cR(\eta,\vx)}$ as
\begin{eqnarray}
\braket{\cR(\eta,\vx)} &\simeq & \f{-H\eta}{\sqrt{2\epsilon_1}}
\int \f{\d^3\vk}{(2\pi)^{3/2}} \f{e^{i \vk \cdot \vx}}{\sqrt{2k}}
[C(\vk)e^{-ik\eta} \nn \\ 
& & + C^\ast(-\vk)e^{ik\eta}]\,.
\label{eq:Rx_subH}
\end{eqnarray}
The rapidly oscillating terms $e^{\pm i k \eta}$ ensure that the integral over $k$ shall be negligible in this regime, as long as $C(\vk)$ does not contain any pathologies.

\paragraph{Two-point function:}
The two-point function of the $\cRk$ in the coherent
state is computed to be
\begin{eqnarray}
\bra{C} \cRk \cR_{\vk'} \ket{C} 
&=& \frac{1}{z^2}\big[ \abs{v_k}^2  \Dd (\bm{k} + \bm{k}') 
+ v_k v_{k'}  C(\vk) C(\vk') \nn \\
& & + v_k v^*_{k'}  C(\bm{k}) C^*( - \bm{k}') \nn \\
& & + v^*_k v_{k'} C^*( - \bm{k}) C(\bm{k}') \nn \\
& & + v^*_k v^*_{k'} C^*( - \bm{k}) C^*( - \bm{k}') \big].
\end{eqnarray}
In the super-Hubble limit, ($|k \eta |, |k' \eta|\to 0$), the products of the
pair of mode functions can be approximated by the terms that are leading order in $\abs{k \eta}^{-1}$. For example,
\begin{align}
  v_k v_{k'}
  \simeq - \frac{1}{2(k k')^{3/2} \eta^2 }.
\end{align}
Using this behavior, the two-point function can be expressed as
\begin{eqnarray}
\label{eq:9927}
\braket{\cRk \cR_{\vk'}} &\simeq &
\frac{1}{z^2} \frac{1}{2 (k k')^{3/2} \eta^2} 
\big[\delta (\bm{k} + \bm{k}') \nn\\
& & - C(\bm{k}) C(\bm{k}') + C(\bm{k}) C^*( - \bm{k}') \nn\\
& & + C^*( - \bm{k}) C(\bm{k}') \nn \\
& & - C^*( - \bm{k}) C^*( - \bm{k}') \big] 
\\
&=& \frac{2 \pi^2}{k^3} \ps(k)  \Dd (\bm{k} + \bm{k}')  \nn\\
& & - 2 \pi^2 \sqrt{\frac{\ps(k) \ps(k')}{k^3 k'^3}} \alpha(\bm{k}) \alpha(\bm{k}').
\end{eqnarray}
Note that, the two-point function has the following structure
\begin{align}
\label{eq:two_p_fun}
\braket{\cRk \cR_{\vk'}}
&= \braket{\cRk \cR_{\vk'}}_{\IR} + \braket{\cRk} \braket{\cR_{\vk'}},
\end{align}
where the first term can be identified as the statistically homogeneous and isotropic irreducible part
\begin{align}
  \braket{\cRk \cR_{\vk'}}_{\IR} = \frac{2 \pi^2}{k^3}
  \ps(k) \Dd (\bm{k} + \bm{k}'),
\end{align}
and the second term is the reducible contribution, entirely due to the non-zero one-point function. 

\paragraph{Three-point function:}
In a similar fashion, the three-point function of $\cRk$ can be evaluated in the coherent state. We ignore the contribution to this correlation from higher-order interactions of the theory as stated in the main text. 
When the modes involved are in the super-Hubble regime, such a three-point function is computed to be 
\begin{align}
\label{eq:9925}
\bra{C} \cR_{\vk_1}\cR_{\vk_2}\cR_{\vk_3} \ket{C}
\simeq & \f{1}{z^3}\bigg[-\frac{i \alpha(\bm{k}_3)}{\left( 2 k_1 k_2 k_3 \right)^{3/2} \eta^3}
\Dd(\bm{k}_1 + \bm{k}_2) \nn\\
& + \text{2 permutations}\bigg] \nn\\
&+ \f{i}{z^3} \frac{\alpha(\bm{k}_1) \alpha(\bm{k}_2) \alpha(\bm{k}_3)}{\left( 2 k_1 k_2 k_3 \right)^{3/2} \eta^3}\,,
\end{align}
which can be expressed in a compact form as
\begin{align}
\label{eq:three_p_fun}
\Braket{\cR_{\vk_1} \cR_{\vk_2} \cR_{\vk_3}}
= & \big[\braket{\cR_{\vk_3}} \braket{\cR_{\vk_1}\cR_{\vk_2}}_{\IR} \nn \\
& + \text{2 permutations}\big] \nn \\
& + \braket{\cR_{\vk_1}} \braket{\cR_{\vk_2}} \braket{\cR_{\vk_3}}\,.
\end{align}
We obtain the last line by comparing with the one- and two-point functions from Eqs.~\eqref{eq:one_p_fun} and~\eqref{eq:two_p_fun}. 

\subsection{Angular correlations in CMB}
The information about the initial quantum state is encoded in the
statistical properties of the CMB. 
Here, we present the angular correlations in CMB due to primordial perturbations arising from a coherent state. 
We utilize [cf.~Eqs.~\eqref{eq:one_p_fun},~\eqref{eq:two_p_fun} and~\eqref{eq:three_p_fun}] to evaluate these correlations.

The line-of-sight temperature fluctuation of a photon, freestreaming
from a direction $\hat{n}$ from the last scattering surface, is expanded in the spherical harmonics as usual
\begin{align}
  \Theta(\hat{n}) = \sum_{\ell = 0}^{\infty} \sum_{m = -\ell}^{\ell} a_{\ell m} Y_{\ell m}(\hat n).
\end{align}
The angular coefficients $a_{\ell m}$  can be related to primordial fluctuations as (for details see e.g.~\cite{baumann2022})
\begin{align}
  a_{\ell m} = 4 \pi i^\ell \int \frac{\d ^3 \bm{k}}{(2 \pi)^{3/2}} \Theta_\ell(k) \cRk Y^*_{\ell m} (\hat {\bm{k}}),
\end{align}
where $\cRk$ is the primordial curvature perturbation and $\Theta_\ell$ is the transfer function (for details of the transfer function see e.g. \cite{dodelson2021}).
Setting the ensemble average of the curvature perturbation to be
$\braket{\cRk} = \bra{C}\cRk\ket{C}$ from Eq.~\eqref{eq:one_p_fun}, the ensemble average of the angular coefficients becomes
\begin{align}
\label{eq:9912}
 \Braket{a_{\ell m}}
 &= 2^{5/2} \pi^2 i^{\ell + 1} \int \frac{\d^3 \bm{k}}{(2 \pi)^{3/2}} \Theta_\ell(k) \sqrt{\ps(k)}
    \frac{\alpha(\bm{k})}{k^{3/2}} Y^*_{\ell m}(\hat{k}) \nonumber\\
 &\equiv  f_{\ell m},
\end{align}
where we define the non-zero average of the angular coefficients as
$f_{\ell m}$. Further, using the decomposition of the form
$\alpha(\bm{k}) = -i \bar\alpha(k) \beta(\hat{\vk})$, the average
can be factorized as
\begin{subequations}
\begin{align}
f_{\ell m} &= f_\ell \beta_{\ell m},\quad \text{with} \\
f_l &\equiv  2^{5/2} \pi^2 i^\ell \int \frac{\d {k}}{(2 \pi)^{3/2}} k^2 \Theta_\ell(k) \sqrt{\ps(k)} \bar{\alpha}(k),\\
    \beta_{\ell m} &\equiv \int \d \Omega_{\hat{k}} \beta(\hat{\vk})Y^*_{\ell m}(\hat{\vk}).
\end{align}
\end{subequations}
We see that for a constant $\beta(\hat{\vk})$, the angular one-point
function vanishes, $f_{\ell m} = 0$ (except when $\ell,m=0$). This
occurs if the primordial mean in the coherent state is independent
of the direction of the wavevector, i.e. $\alpha(\bm{k}) = \alpha(k)$.

\paragraph{Angular two-point function.}
The angular two-point function of the CMB perturbation can be
written as
\begin{eqnarray}
\Braket{a_{\ell m} a_{\ell' m'}} 
&=& (4 \pi)^2 i^{\ell+\ell'} \int \frac{\d^3 \bm{k}}{(2 \pi)^{3/2}} \int \frac{\d^3 \bm{k}'}{(2 \pi)^{3/2}} \Theta_\ell(k) \Theta_{\ell'}(k') \nn\\
& & \times Y^*_{\ell m}(\hat k) Y^*_{\ell' m'} (\hat{k'}) 
\braket{\cRk \cR_{\vk'}}.
\end{eqnarray}
Taking the two-point function of the primordial curvature perturbation to be
$\braket{\cRk\cR_{\vk'}} =\bra{C} \cRk \cR_{\vk'} \ket{C}$ 
and using Eq.~\eqref{eq:two_p_fun}, the above expression becomes
\begin{align}
  \Braket{a_{\ell m} a_{\ell' m'}} =
  \Braket{a_{\ell m} a_{\ell' m'}}_{\IR}
  + f_{\ell m} f_{\ell' m'}.
\end{align}
The first term is the statistically homogeneous and isotropic
irreducible part that takes the form of the usual angular power
spectrum,
\begin{align}
  \label{eq:995}
  \Braket{a_{\ell m} a_{\ell' m'}}_{\IR} &=  C_\ell \delta_{\ell \ell'} \delta_{m m'},\quad \text{where}\\
  C_\ell &=  4 \pi \int \d \ln k  \ps(k) \Theta_\ell^2(k).
\end{align}
The statistically inhomogeneous and anisotropic additional term is
solely sourced by the angular one-point function. Similar
to the case of the primordial correlation functions, the angular
two-point function decomposes into its irreducible and
inhomogeneous-anisotropic parts.
Note that, if the primordial mean is independent of the direction of the wavevector 
i.e. it is inhomogeneous but isotropic, with $\alpha(\bm{k}) = \alpha({k})$, then
$f_{\ell m} = 0$ (for $\ell,m \neq 0$) and so the corresponding contribution to the angular two-point function vanishes.

\paragraph{Angular three-point function.}
The angular three-point function can be written in terms of the
primordial curvature perturbation as
\begin{align}
  &\Braket{a_{\ell_1 m_1} a_{\ell_2 m_2} a_{\ell_3 m_3} } \nonumber\\
  =& (4 \pi)^3 i^{(\ell_1 + \ell_2 + \ell_3)} \iiint \frac{\d^3 \bm{k}_1}{(2 \pi)^{3/2}} \frac{\d^3 \bm{k}_2}{(2 \pi)^{3/2}} \frac{\d^3 \bm{k}_3}{(2 \pi)^{3/2}} \nonumber\\
  &\times \Theta_{\ell_1}(k_1) \Theta_{\ell_2}(k_2) \Theta_{\ell_3}(k_3)
    Y^*_{\ell_1 m_1}(\hat{k}_1) Y^*_{\ell_2 m_2}(\hat{k}_2) Y^*_{\ell_3 m_3}(\hat{k}_3) \nonumber\\
  &\times \Braket{\cR_{\vk_1}\cR_{\vk_2}\cR_{\vk_3}}.
\end{align}
Setting the ensemble average of 
$\Braket{\cR_{\vk_1}\cR_{\vk_2}\cR_{\vk_3}} = \bra{C}\cR_{\vk_1}\cR_{\vk_2}\cR_{\vk_3}\ket{C}$ and using Eq.~\eqref{eq:three_p_fun} in the above expression we get
\begin{subequations}
  \begin{align}
    &\Braket{a_{\ell_1 m_1} a_{\ell_2 m_2} a_{\ell_3 m_3} } \nn \\
    =& \left\{ \Braket{a_{\ell_3 m_3}}\Braket{a_{\ell_1 m_1} a_{\ell_2 m_2}} +
       \text{2 permutations} \right\} \nonumber\\
    &+ \Braket{a_{\ell_1 m_1}}\Braket{a_{\ell_2 m_2}}\Braket{a_{\ell_3 m_3} } \\
    =& \left\{ f_{\ell_3 m_3} C_{\ell_1} \delta_{\ell_1 \ell_2} \delta_{m_1 m_2}
       + \text{2 permutations} \right\} \nonumber\\
    &+ f_{\ell_1 m_1} f_{\ell_2 m_2} f_{\ell_3 m_3},
  \end{align}
\end{subequations}
where we have used the angular one- and two-point functions
[Eqs.~\eqref{eq:9912} and~\eqref{eq:995}] to obtain the last line. 
As expected in a Gaussian theory, the three-point function does not contain an irreducible part $\Braket{a_{\ell_1 m_1} a_{\ell_2 m_2} a_{\ell_3 m_3} }_{\IR}$. 
Since it exits entirely due to the reducible parts, the three-point function
is non-zero only when the primordial fluctuations have a non-zero
mean. It vanishes when the mean is isotropic. 

\section{A note on squeezed and squeezed coherent states}\label{sup:sq-coh}

Among the non-Bunch-Davies vacuum states, squeezed states are widely considered for primordial perturbations~\cite{Holman:2007na,Aravind:2013lra,Kundu:2013gha,Chen_2014,Ragavendra:2020vud,Kanno:2022mkx,Akama:2023jsb}.
There are some works that consider squeezed coherent states as initial states for primordial perturbations~\cite{Koh:2004ez,Dona:2016ohf,Mondal:2024glo}.
In this section, we clarify that our results are specific to initial states being coherent states, not squeezed states and squeezed coherent states can be regarded as extension to our analysis.

The creation and annihilation operators of the squeezed state, say $\ket{S}$, (let us call them $\tilde a^\dagger_\vk$ and $\tilde a_\vk$) are related to those of the Bunch-Davies vacuum through the Bogoliubov transformation (for details, see Ch.~5 of Ref.~\cite{Loudon:2000})
\begin{eqnarray}
\tilde{a}_{\vk} &=& A_k a_\vk + B_k^\ast a_{-\vk}^\dagger\,, \\
\tilde{a}^\dagger_{\vk} &=& A_k^\ast a^\dagger_\vk + B_k a_{-\vk}\,,
\end{eqnarray}
where $A_k$ and $B_k$ are the Bogoliubov coefficients.
This implies that evaluating expectation values of operators involving $\cR$ in squeezed state is equivalent to them being evaluated in Bunch-Davies vacuum, but with modified mode function $\tilde f_k$, that are related to Bunch-Davies mode function $f_k$ as
\begin{eqnarray}
\tilde f_k &=& A_kf_k + B_kf_k^\ast\,.
\end{eqnarray}
The constraint on the Wronskian of $f_k$ translates to $\vert A_k \vert^2-\vert B_k \vert^2=1$.
The transformation mixes the positive and negative frequency parts of the Bunch-Davies mode function and setting $B_k=0,\, A_k=1$ restores Bunch-Davies vacuum.
The expectation value of $\cRk$ in the squeezed state $\ket{S}$ shall be 
\begin{eqnarray}
\bra{S} \cRk \ket{S}=0, 
\label{eq:1ptsq}
\end{eqnarray}
as in the Bunch-Davies vacuum.
However, the higher-order correlations of $\cRk$ shall be modified by $A_k$ and $B_k$ and they have been extensively studied (for e.g.~\cite{Chen_2007,Meerburg2009,Agullo:2010ws,Meerburg:2010rp,Kanno:2022mkx,Shiraishi_2013,Chen_2014,Chen_2015,Akama:2020jko}).
For instance, the two-point correlation in the super-Hubble regime shall be~\cite{Ragavendra:2020vud}
\begin{eqnarray}
\bra{S} \cR_{\vk_1} \cR_{\vk_2}\ket{S}
&=& \f{2\pi^2}{k^3} \ps(k) \vert A_k - B_k \vert^2 \delta^{3}(\vk_1+\vk_2)\,.
\label{eq:2ptsq}
\end{eqnarray}
Evidently, the squeezed initial state, while altering the shape of correlations, does not violate statistical homogeneity or isotropy of the perturbations.

In the squeezed coherent state (let us denote it as $\ket{C,S}$), the creation and annihilation operators (say $\tilde{\tilde a}^\dagger_\vk$ and $\tilde{\tilde a}_\vk$) are related to the corresponding operators of the vacuum as~\cite{Loudon:2000}
\begin{eqnarray}
\tilde{\tilde{a}}_{\vk} &=& \tilde a_\vk + C(\vk)
= A_k a_\vk + B_k^\ast a_{-\vk}^\dagger + C(\vk)\,, \\
\tilde{\tilde{a}}^\dagger_{\vk}  &=& \tilde a_\vk + C^\ast(\vk) 
= A_k^\ast a^\dagger_\vk + B_k a_{-\vk} + C^\ast(\vk)\,,
\end{eqnarray}
where the coherent state essentially displaces the operators with the function $C(\vk)$.
In such a state the expectation value of $\cRk$ becomes
\begin{eqnarray}
\bra{C,S}\cRk \ket{C,S} &=& C(\vk)\tilde{f}_k+C^\ast(-\vk)\tilde{f}^\ast_k\,.
\end{eqnarray}
We may then use the behavior of $f_k(\eta)$ in super-Hubble regime to reduce it to
\begin{eqnarray}
\bra{C,S}\cRk \ket{C,S} & \simeq & -i\sqrt{\frac{2 \pi^2}{k^3}\ps(k)}
\bigg[C(\vk) (A_k - B_k) \nn \\
& & - C^*(-\vk) (A^\ast_k - B^\ast_k) \bigg]\,.
\end{eqnarray}
Clearly, the coherent state induces a non-zero value for the one-point function while the effect of squeezing is captured in the factors of $A_k-B_k$. 
The higher-order correlations in this state shall have shapes altered by $A_k$ and $B_k$, along with reducible contributions containing factors of $C(\vk)$.

If we set $B_k=0$ (and so $A_k=1$), we recover our results due to coherent state, as discussed in the main text.
If we retain $B_k \neq 0$, but set $C(\vk)=0$, we recover the results of a squeezed state seen in Eqs.~\eqref{eq:1ptsq} and~\eqref{eq:2ptsq}.
In our work, we do not invoke squeezing of the coherent state $\ket{C}$ and focus on the non-vanishing behavior of $\bra{C}\cRk\ket{C}$. 
Of course, one can extend this analysis invoking squeezing of the coherent state, without demanding $\bra{C,S}\cRk\ket{C,S}=0$ and explore additional features introduced by non-zero values of $B_k$ along with $C(\vk)$ in the N-point correlations.

\section{Energy density due to statistically inhomogeneous fluctuations}

Evolving the primordial perturbations from coherent state contributes to additional energy density in perturbations.
We must ensure that the choice of initial state does not lead to backreaction nor introduce any divergences in the system. 
In this section, we compute the inhomogeneous contribution to the perturbation energy density due to initial coherent state and discuss the associated issues.

Under the slow-roll approximation, the energy density operator associated with the curvature perturbation can be written as~\cite{Abramo_1997} (also see~\cite{Kundu:2011sg} in the context of coherent state)
\begin{align}
{\rho}_{\mathcal{R}} \approx \frac{z^2}{a^4} \left( \frac{1}{2} \cR'^2  + \frac{1}{2} \left( \nabla \cR \right)^2 \right).
\end{align}
The expectation value of this energy density evaluated in the Bunch-Davies vacuum takes the form
\begin{align}
\Braket{\rho_{\cR}}_0
&= \frac{z^2}{2 a^4}   
\int \frac{\d^3 \bm{k}}{(2 \pi)^3} \left( \abs{f'_k}^2
+ {k^2}   \abs{f_k}^2 \right),
\end{align}
where $f_k(\eta) = v_k(\eta)/z$ and the subscript $0$ denotes evaluation in the vacuum state. 
During the early stages of inflation ($\eta \to - \infty$), when all relevant modes are inside the sub-Hubble regime, the energy density can be approximated as
\begin{align}
\label{eq:rho_v}
\Braket{\rho_{\mathcal{R}}}_0
&\approx \frac{1}{4 \pi^2 a^4} \int \d k k^3 + \frac{H^2}{8 \pi^2 a^2} 
\int \d k k.
\end{align}
Therefore, the energy density varies as $a^{-4}$ at the early times,
which is an expected behaviour for a massless scalar field. 
The UV divergence appearing in the energy density is renormalized using a regularization scheme (see, for example,~\cite{Allen_1987,Bunch_1980,Parker_1974,Kundu:2011sg}).
The renormalized value of this energy density is shown to be 
$\braket{\rho_\cR}_0 = 61H^4/(960\pi^4)$~\cite{Allen_1987}.
This value is clearly sub-dominant to the background energy density during inflation $\rho_{\rm I} = 3H^2\Mpl^2$, as $H$ is typically $H \simeq 10^{-5}\,\Mpl$.

As to the expectation value of $\rho_\cR$ evaluated in the coherent state, we can show that 
\begin{align}
  \bra{C}{\rho}_{\mathcal{R}} \ket{C}
  =  \Braket{{\rho}_{\mathcal{R}}}_0
  +   \Braket{{\rho}_{\mathcal{R}}}_{\text{IA}},
\end{align}
where the first term is the homogeneous-isotropic contribution, same
as the energy density in the Bunch-Davies vacuum state. The second
term is the additional contribution from the statistically
inhomogeneous-anisotropic part, which is of the form
\begin{align}
  \Braket{{\rho}_{\mathcal{R}}}_{\text{IA}}
  &= \frac{z^2}{a^4} \left( \frac{1}{2} \Braket{\cR'^2}_{\text{IA}}  +
    \frac{1}{2} \Braket{ \left( \nabla \cR \right)^2}_{\text{IA}} \right)\\
  &= \frac{z^2}{2 a^4}\left( \tilde{R}' + \tilde{R}'^* \right)^2 -
    \frac{1}{2 a^4} \left( \tilde{\bm{u}} - \tilde{\bm{u}}^* \right)^2,
\end{align}
where we have defined 
\begin{align}
  \tilde{R}'(\eta, \bm{x})
  &= \int \frac{\d^3 \bm{k}}{(2 \pi)^{3/2}} \e^{i \bm{k} \cdot \bm{x}} C(\bm{k}) f_k'(\eta)\,, \\
  \tilde{\bm{u}}(\eta, \bm{x})
  &= \int \frac{\d^3 \bm{k}}{(2 \pi)^{3/2}} \bm{k} \e^{i \bm{k} \cdot \bm{x}} C(\bm{k}) v_k(\eta).
\end{align}

Since the overall scaling is as $a^{-4}$, we focus on the behavior of this quantity at early times, when the relevant modes are in the sub-Hubble regime.
When $\eta \to -\infty$, the energy density in the inhomogeneous-anisotropic part can be approximated as
\begin{align}
  \Braket{{\rho}_{\mathcal{R}}}_{\text{IA}}
  \approx
  &\frac{1}{4 a^4} \Bigg\{\left[ \int \frac{\d k \, \d^2\Omega_{\hat \vk}}{(2 \pi)^{3/2}}  \,\e^{i \bm{k} \cdot \bm{x}} \, k^{5/2} \right. \left( i\,C(\bm{k})  {\e^{-i k \eta}} \right.\nonumber\\
  &\qquad \left.\left. -  i\,C^*(-\bm{k})  {\e^{i k \eta}}  \right)\right]^2 \nonumber\\
  &- \int \int \frac{\d k_1 \, \d^2\Omega_{\hat \vk_1}}{(2 \pi)^{3/2}} 
  \frac{\d k_2 \, \d^2\Omega_{\hat \vk_2}}{(2 \pi)^{3/2}} \,\e^{i (\vk_1+\vk_2) \cdot \bm{x}} \, (\hat{\vk}_1 \cdot \hat{\vk}_2) \nn \\
  &\times (k_1k_2)^{5/2} \left(C(\vk_1) \e^{-i k_1 \eta} +  C^\ast(-\vk_1) \e^{i k_1 \eta}\right) \nn \\
  & \times \left( C(\vk_2) \e^{-i k_2 \eta} + C^*(-\vk_2) \e^{i k_2 \eta} \right)\Bigg\}.
\end{align}
Unlike the vacuum energy density $\Braket{{\rho}_{\mathcal{R}}}_0$, the additional energy density above explicitly depends on position, due to the breaking of statistical homogeneity and isotropy. 
Moreover, the integrands are highly oscillatory due to $\exp(\pm i k\eta)$ terms, which vanish upon integration. 
So, with reasonably well-behaved $C(\vk)$, at early times when modes are in the sub-Hubble regime 
($\vert k\eta\vert \gg 1$), $\braket{\rho_\cR}_{\rm IA}$ leads to negligible contribution.
Besides, we can see from the above equation that, if $C(\vk)$ decays faster than $k^{-5/2}$ for large values of $k$, it does not lead to any new divergences in the UV limit either (as already noted in Ref.~\cite{Kundu:2011sg}).

The behavior of $\braket{\rho_\cR}_{\rm IA}$ is similar to the vanishing of $\braket{\cR(\eta,\vx)}$ as $\eta \to -\infty$ [Eq.~\eqref{eq:Rx_subH}].
Therefore, the additional energy density $\braket{\rho_\cR}_{\rm IA}$ due to coherent states, is negligible compared to $\braket{\rho_\cR}_0$ and does not cause any backreaction.

\section*{Other Implications}\label{sec:SM2}

The breakdown of statistical homogeneity and isotropy have other implications, which we discuss here.

In this context, it is pertinent to note three different notions related to measuring the galaxy power spectrum/CMB angular power spectrum. The first relates
to average over initial quantum ensembles in a coherent state, which we have used to determine the primordial correlation functions.
The second relates to average over classical ensembles which might be  relevant during the super-Hubble evolution of modes (for a discussion on the transition between these two averages, see~\cite{weinberg2008}). 
The third method is implicit in defining  the estimator in Eq.~(12) of the manuscript. 
Here the sum over modes for a fixed $|\vk|$ is used as a proxy for ensemble
average and it follows from ergodic hypothesis. 

The ergodic hypothesis is violated if statistical homogeneity doesn't hold (e.g. Appendix~D of \cite{weinberg2008}).  This doesn't allow us to relate ensemble average to spatial averages, which 
is key to analyzing cosmological data. For instance, the  power spectrum estimator in Eq.~(12) of the manuscript is unbiased for a statistical homogeneous density field because each Fourier mode
for a fixed $|k|$ is uncorrelated  and therefore the average over such Fourier modes is equivalent 
to ensemble averaging. This also allows us to reduce error on the estimated power spectrum as
Eq.~(15) in the manuscript shows. As the ergodic hypothesis is violated in our case, the estimator given by Eq.~(12) in the manuscript might only yield a partial picture.

Therefore,  the analysis of data corresponding to a  maximally inhomogeneous density field would require major rethinking. However, we show that  the current data constrains this component to be small. One of our focus here is to seek unique signatures of this violation, e.g. cross-correlating galaxy data for $\vk_1 \ne \vk_1$, potential detection of bispectrum of amplitude and shape
not expected in the usual theory [Eq.~(23) in the manuscript], anisotropy of $\mu$-distortion, etc.

Based on our discussion, the density field is expected to be mix of statistically homogeneous and inhomogeneous components, with the latter component being subdominant. This also  alters theoretical modelling
of the system.

The cosmological perturbation theory is based on splitting general
space and time-dependent quantities into a time-dependent and
space-independent background (FLRW universe for maximally symmetric
spaces) and time and space-dependent fluctuations. The only
condition on the space-dependent quantities is that they be
`small', e.g. $\braket{\cR(\vx)} \ll 1$. This can be ensured in a
gauge-invariant way for both metric and matter variables (e.g. for details see \cite{dodelson2021} and references therein). Therefore, Eq.~(\ref{eq:one_p_fun}) doesn't impact
the theoretical framework of general relativistic perturbation
theory so long as the perturbed quantities remain small. 

In the standard cosmological theory based on
linear, Gaussian perturbations, the power spectrum given of a statistically homogeneous process (the first term of  Eq.~\eqref{eq:two_p_fun}) provides the seed for perturbations, e.g.
for adiabatic initial conditions.  The curvature perturbation  for a
given superhorizon scale $k$ acquires a value equal to the square
root of the power spectrum and the fluctuations of
other metric and matter variables can be related to  this variable. The correlation properties of these quantities, e.g.
density and CMB temperature fluctuations, are then naturally 
inherited from the assumption of  statistical homogeneity.  When the second term on the RHS
of Eq.~(\ref{eq:two_p_fun}) is present, the initial conditions are altered. 
In this case, the initial conditions are a sum of two uncorrelated random variables, one with zero mean and the other proportional to $\alpha(\vk)$. 

\bibliographystyle{apsrev4-2}
\bibliography{lib}

\end{document}